\documentclass[prl,twocolumn,graphicx,amssymb,floatfix]{revtex4}
\usepackage{graphicx}
\usepackage{amsmath}
\usepackage{amssymb}
\usepackage{amsthm}
\usepackage{color}
\usepackage{tikz}
\usetikzlibrary{calc,spy}

\begin{document}

\title{From  Entanglement Witness to Generalized Catalan Numbers}
\author{E. Cohen$^{1,2}$, T. Hansen$^{3,4}$, N. Itzhaki$^2$}
\affiliation{$^1$H.H. Wills Physics Laboratory, University of Bristol, Tyndall Avenue, Bristol, BS8 1TL, U.K\\
             $^2$School of Physics and Astronomy, Tel-Aviv University, Tel-Aviv 6997801, Israel\\
             $^3$Centro de F\'\i sica do Porto, Faculdade de Ci\^encias da Universidade do Porto,
                 Rua do Campo Alegre 687, 4169--007 Porto, Portugal\\
             $^4$II.\ Institut f\"ur Theoretische Physik, Universit\"at Hamburg, Luruper Chaussee 149, D-22761 Hamburg, Germany}

\begin{abstract}
The problem of entanglement detection for arbitrary spin systems is analyzed. We demonstrate how a single measurement of the squared total spin can probabilistically discern separable from entangled many-particle states. For achieving this goal, we construct a tripartite analogy between the degeneracy of entanglement witness eigenstates, tensor products of $SO(3)$ representations and classical lattice walks with special constraints. Within this framework, degeneracies are naturally given by generalized Catalan numbers and determine the fraction of states that are decidedly entangled and also known to be somewhat protected against decoherence. In addition, we introduce the concept of a ``sterile entanglement witness'', which for large enough systems detects entanglement without affecting much the system's state. We discuss when our proposed entanglement witness can be regarded as a sterile one.

\end{abstract}
\maketitle

\section{Introduction}

It is clear by now, that the phenomenon of quantum entanglement lies at the heart of quantum mechanics. Entanglement is recognized as an important resource for quantum computation \cite{Com}, quantum cryptography \cite{Cry}, quantum teleportation \cite{Tel}, quantum black holes \cite{Harlow} and many other quantum tasks. It was also demonstrated experimentally that entanglement can affect macroscopic properties of solids, albeit at very low (critical) temperature (below 1 Kelvin) \cite{Bru}.

Any quantum state $\rho$, has an ensemble decomposition, i.e. there exist quantum states $\rho_i$  with a probability
distribution $p_i$ such that $\rho=\sum p_i \rho_i$. If there exists an ensemble decomposition where every $\rho_i$ is a separable state, then $\rho$ is called a separable state; otherwise, it is called an entangled state.

Detecting entanglement of a given state, however, is known to be a hard computational problem (NP) \cite{Gur}. Several methods of detecting entanglement are Bell and spin squeezing inequalities \cite{Ter,Kit}, measurement of nonlinear properties of the state \cite{Gun}, approximation of positive maps \cite{Kor}. We shall focus in this paper on a method known as {\it entanglement witnesses} \cite{Hor}. This method is unique because it is valid for any quantum system, regardless of the number and dimensions of its subsystems \cite{Wu}. An entanglement witness (EW) is an Hermitian (non-positive) operator, whose expectation value is positive for any separable state. Therefore, when applied to a state of interest, a negative expectation value directly indicates the entanglement in this state. Based on the theorem below \cite{Hor}, the EW is a necessary and sufficient condition for entangled states: \\

Theorem (Horodecki96) :  A density matrix $\rho$ on $H_A\otimes H_B$ is entangled
if and only if there exists a Hermitian operator $W$, an {\it entanglement witness}, such that
\begin{equation}
TrW\rho<0,
\end{equation}
and for all separable states $\rho_{sep}$,
\begin{equation}
TrW\rho_{sep} \ge 0.
\end{equation}

Furthermore, for every entangled state, there exists an EW to detect it \cite{Hor}. However, a specific EW can have a positive expectation value also when evaluated on entangled states. In fact, no entanglement witness can discern separable states from entangled ones with 100\% success rate. That is, there are always undecideable states with respect to any entanglement witness (hopefully, not too many, because this will render the EW ineffective). We do not consider in this work the optimality property of the EW, but for the sake of completeness we note that an entanglement witness is said to be optimal if there exists no other EW which is finer, i.e. has a larger set of decideable states \cite{Lewe}.

A drawback of this EW method is that we change the system's state when measuring the witness (unless the system is in some particular eigenstate). We shall further analyze this feature and see how to overcome it in Sec.\ 1 using what we term a ``sterile'' entanglement witness. We will show that for many decideable states, one can evaluate the witness while negligibly changing the state of the system.

We will focus on spin systems having nearest-neighbors interactions, with an EW corresponding to a Heisenberg model without an external magnetic field. This EW describes fully coupled $N$ spin $s$ particles in the form of a complete graph \cite{Plesch}.

The outline of the work is as follows. We begin by presenting in Sec.\ 1 the proposed EW and show it is a ``sterile'' one. In Sec.\ 2, a tripartite analogy is discussed between degeneracies of the witness eigenstates, tensor products of $SO(3)$ representations and lattice paths which generalize the Catalan numbers. A few examples are analyzed. We then derive the fraction of decideable states for various $s$ and $N$ values in Sec.\ 3 and show that it remains finite when $N\to\infty$.

\section{1. A Sterile Entanglement Witness}

The proposed EW is given by the square of the total spin operator, or Casimir operator
\begin{equation} \label{H}
W=\mathbf{J}^2 = J_x^2+J_y^2+J_z^2,
\end{equation}
where $J_{x/y/z}=\sum_{k=1}^N s_{k,x/y/z}$ is the total spin in each direction.
The eigenvalues $W_j$ are given by the familiar eigenvalues of $\mathbf{J}^2$
\begin{equation} \label{Wj}
W_j = j (j+1), \quad j \in \{\tfrac{n}{2}, n \in \mathbb{N}_0 \}.
\end{equation}
This EW was previously studied by T$\acute{\text{o}}$th \cite{Toth}, who used the above model for the case of spin $s = \frac{1}{2}$ particles.

The existence of undecideable states can be most easily demonstrated in a simple system of two spins. The states $|\uparrow\downarrow\rangle$ and $|\downarrow\uparrow\rangle$ are both separable. The proposed EW is a linear operator, hence we may consider their superposition. On the one hand, these states construct the spin-0 singlet state $\frac{1}{\sqrt{2}}(|\uparrow\downarrow\rangle-|\downarrow\uparrow\rangle)$, which is maximally entangled. It has the lowest possible $W$ value, and hence will be identified by it. On the other hand, a different superposition $\frac{1}{\sqrt{2}}(|\uparrow\downarrow\rangle+|\downarrow\uparrow\rangle)$, having spin $1$, will not be recognized as an entangled state
since $|\uparrow\uparrow\rangle$ and $|\downarrow\downarrow\rangle$
have $j=1$ too.

For a spin $\frac{1}{2}$ system, the spin operator is given in terms of the three Pauli matrices
$J_{x/y/z}=\sum_{k=1}^N \frac{1}{2} \sigma_{k,x/y/z}$.
The expectation value of separable spin 1/2 states is bounded
\begin{equation} \label{boundhlaf}
\langle W \rangle \ge W_{sep}^{(1/2)}=N/2.
\end{equation}
Hence, if the measured value of the EW is small enough, the $N$-particle state is understood to be entangled, while if the EW is high, we cannot tell with certainty if the state is entangled or separable.
However, by knowing the degeneracy of the witness eigenstates, we can determine the fraction of all states which are decideable.

The degeneracy of states with eigenvalue $W_j$ for even $N$ were analytically found in \cite{Toth,Cirac}
\begin{equation} \label{dj}
d_{\frac{1}{2}}(N,j)=\begin{cases}
\frac{(2j+1)^2}{N/2+j+1} \dbinom{N}{N/2+j}, &j \in \mathbb{N}_0,\\
0, &\text{else}.
\end{cases}
\end{equation}
As will be described in Sec.\ 2, $d_{\frac{1}{2}}(N,j)$ are strictly related to the so-called Catalan triangle \cite{Sta}. Moreover, we will derive in Sec.\ 2 the above formula from the structure of $SO(3)$ tensor products and generalize it to systems of arbitrary spins. We will also relate this problem to a classical problem of enumerative combinatorics - finding the number of constrained lattice paths in 2D.

The question now is whether we can detect entanglement using the above operator $W$ without disturbing much the local dynamics of the system given by some Hamiltonian $H_L$. In other words, we would like to verify that
\begin{equation} \label{weak}
[H_L, W] = o(N),
\end{equation}
when evaluated in some subspace of entangled states, i.e. the commutation relation, being a sub-extensive quantity, is asymptotically dominated by the size of the system. This is, of course, not the usual notion of commutation (which evidently is not satisfied by our EW), but we find it more appropriate for describing weak operations on large systems as will be shown below.

To demonstrate \eqref{weak} we shall use the quite general nearest-neighbors interactions within a 3D homogenous Heisenberg lattice:
\begin{equation}
H_L = \frac{1}{2} J \sum_{k=1}^N {\bf s}^k {\bf s}^{k+1},
\end{equation}
where ${\bf s}^k$ is the vector $(s_x, s_y, s_z)$ of the $k$th particle, and the coupling constant $J>0$ (corresponding to the anti-ferromagnetic case) is not necessarily small.

Using the well-known commutation relations between angular momentum operators, we find:
\begin{equation} \label{sweak}
[H_L, W] = iJ\sum_{i=1}^N \{ [J_x, s_y^is_z^{i+1}]+[J_y, s_z^is_x^{i+1}]+[J_z, s_x^i s_y^{i+1}]\}.
\end{equation}

Anti-parallel spins are obviously preferred by this local Hamiltonian. When the EW is evaluated, for instance, in the ground state of $H_L$, we have $J_{x/y/z} =O(\sqrt{N})$ (the total $J$ along each axis is low because only a few spins do not cancel), and hence \eqref{weak} follows. This can be easily seen also in the first eigenstates of $H_L$, where both $H_L$ and $W$ remain small. Recall that $W_j$ grows as $j^2$, so if $j=o(\sqrt{N})$ and $\langle H_L \rangle=O(1)$, \eqref{weak} will be satisfied. This relation between $H_L$ and $W$ maintains its meaning until $\langle H_L \rangle \approx -3N$, which is the minimal energy of separable states in this Heisenberg lattice model \cite{Toth}. Intuitively, it is clear that when having a large system in one of its lower, highly entangled states, $J_{x/y/z}$ would be negligibly affected if evaluated on this state or this state with two altered spins, as in \eqref{sweak}. This means that an entangled state, as well as the system's energy corresponding to it,  are likely to change only slightly after the entanglement witness $W$ has been applied. The same does not hold though, for a separable state. In the limit of $N\rightarrow\infty$, the asymptotic relation in \eqref{weak} implies an approximate notion (in the macroscopic sense) of commutation. In this limit, the fraction of non-parallel spins goes to zero, while the fraction of entangled states that are decideable stays finite, as we shall see in Sec.\ 3. Hence, the relative change in energy due to the application of the EW gets very small.

These arguments can be straightforwardly applied also to other forms of local dynamics such as the XY model and the Bose-Hubbard model using the suitable bounds which were calculated in \cite{Toth}.

\section{2. The Tripartite Analogy}\label{Tri}
Eq. \eqref{dj} follows from a general relation between tensor products of $SO(3)$ representations and degeneracies of witness eigenstates.
This relation is based on the fact that the eigenvalue $W_j$ only depends on the overall spin $j$ of the eigenstate, which specifies the irreducible representation (irrep) $[j]$ of $SO(3)$ according to which it transforms under rotations
\footnote{These irreps can be constructed explicitly as the vector spaces of traceless symmetric rank $j$ tensors in 3D.}.
As a consequence the degeneracy of states with eigenvalue $W_j$ is given by the number of distinct states of spin $j$.
The number of states of spin $j$ that can be created from $N$ spin $s$ irreps is given by the multiplicity of the irrep $[j]$
in the tensor product
\begin{equation}
[s]^{\otimes N} = \underbrace{[s]\otimes [s] \otimes \ldots \otimes [s]}_{N \text{~times}},
\end{equation}
multiplied by the dimension of this irrep $\text{dim}([j])=2j+1$
\begin{equation} \label{degeneracy}
d_{s}(N,j) = \text{dim}([j]) \ \text{mult}_{[j]} ([s]^{\otimes N}).
\end{equation}
The direct sum decomposition of the $SO(3)$ tensor product can be formulated conveniently in terms of tensor product coefficients $b_{j_1 j_2 j_3}$
\footnote{See e.g.\ \cite{Fuchs:1997jv} for an introduction to the topic.}
\begin{equation}\label{tensor_product}
[j_1] \otimes [j_2] = \bigoplus\limits_{j_3 \in \mathbb{N}_0 / 2} b_{j_1 j_2 j_3} [j_3],
\end{equation}
given by the numbers \cite{Fuchs:1997jv}
\begin{equation}\label{tensor_product_coefficients}
b_{j_1 j_2 j_3} = \begin{cases}
1 , &j_1 + j_2 + j_3 \in \mathbb{N}_0\\
 &\wedge \ |j_1-j_2| \leq j_3 \leq j_1 + j_2,\\
0, &\text{otherwise}.
\end{cases}
\end{equation}
These constraints are familiar from the well-known Clebsch-Gordan coefficients.
The coefficients $b_{j_1 j_2 j_3}$ have the property $b_{\, 0 j_1 j_2} = \delta_{j_1 j_2}$
and are symmetric under any permutation of the three indices,
a consequence of the self-duality of representations of the $SO(d)$ groups.
The multiplicity of a given irrep in a twofold tensor product is trivially given by \eqref{tensor_product}
\begin{equation}
\text{mult}_{[j_3]} ([j_1] \otimes [j_2]) = b_{j_1 j_2 j_3}.
\end{equation}
Similarly the multiplicity in \eqref{degeneracy} is derived by
repeated application of \eqref{tensor_product} on $[s]^{\otimes N}$
\begin{align}\label{multiplicity}
& \text{mult}_{[j]} ([s]^{\otimes N})=\\
&\sum\limits_{j_1,j_2,\ldots,j_{N-1} \in \mathbb{N}_0 / 2}
b_{0 s j_1} b_{j_1 s j_2} \ldots b_{j_{N-2} s j_{N-1}} b_{j_{N-1} s j}.
\nonumber
\end{align}
A connection to the formula \eqref{dj} can be made by visualizing each term in the sum \eqref{multiplicity}
as a path on a 2D lattice connecting the points
\begin{equation}
(0,0) \rightarrow (1,j_1) \rightarrow \ldots \rightarrow (N-1,j_{N-1}) \rightarrow (N,j).
\end{equation}
The paths that contribute to the sum are the ones where every step (say at position $(x,y)$) is according to \eqref{tensor_product_coefficients}
in the set
\begin{equation} \label{lattice_steps}
\left\{
(1,s), (1,s-1), \ldots, (1,\text{max}(-s,s-2y))
\right\}.
\end{equation}
These lattice paths are discussed below with the aid of a few examples
and allow us to obtain recursion relations for the multiplicities $m_s(N,j)\equiv \text{mult}_{[j]} ([s]^{\otimes N})$.
The number of lattice paths up to $y=N$ can be expressed in terms of paths ending
at $y=N-1$
\begin{equation}
m_s(N,j) = \begin{cases}
1, &N=j=0,\\
\sum\limits_{k=|j-s|}^{\text{min}(j+s, (N-1)s)}
m_s(N-1,k), &0 \leq j \leq N s,\\
0, &\text{else},
\end{cases}
\label{eq:m_recursion}
\end{equation}
where $k$ increases in integer steps, i.e.\ it takes only integer or half-integer values depending on whether
$|j-s|$ is an integer or half-integer.
With these recursion relations it is possible to efficiently calculate the multiplicities for high values of $N$.

\subsection{Spin 1/2 particles}

\begin{figure}[htbp]
\begin{center}
\scalebox{0.85}{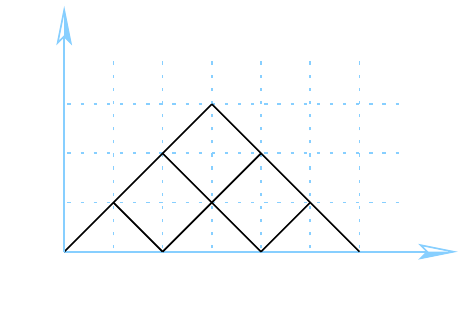}
\quad
\scalebox{0.85}{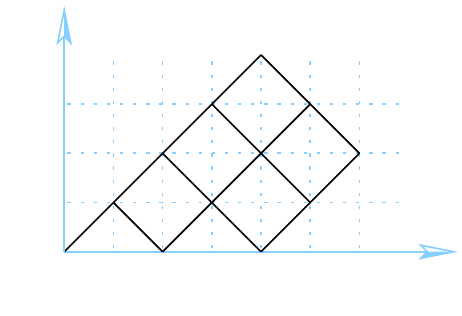}
\caption{Lattice paths in the x-y plane illustrating $\text{mult}_{[0]} ([\frac{1}{2}]^{\otimes 6})=5$ (left) and $\text{mult}_{[1]} ([\frac{1}{2}]^{\otimes 6})=9$ (right).}
\label{fig:spin12ex}
\end{center}
\end{figure}

Let us now give further insight into the lattice paths defined above by discussing the first few examples.
Where available, we will make connections to the mathematical literature regarding the lattice paths.
For $s=\frac{1}{2}$ the paths are given by the allowed steps
\begin{equation}
\begin{aligned}
&\left\{
(1,\tfrac{1}{2}), (1,-\tfrac{1}{2})
\right\}, &y > 0,\\
&\left\{
(1,\tfrac{1}{2})
\right\}, &y = 0.
\end{aligned}
\end{equation}
Two examples are given in Fig.\ \ref{fig:spin12ex}.
These directed lattice paths, which do not go below the $x$-axis and end at a point $(N,j)$ after $N$ steps are known to be counted
by the Catalan triangle $C(N/2+j, N/2-j)$ \cite{Sta}, which is given for any $n\ge k \ge 0$ by
\begin{equation}
C(n,k)=\frac{(n+k)!(n-k+1)}{k!(n+1)!}.
\end{equation}
Using this formula it is easy to see that \eqref{dj} and \eqref{degeneracy}
match.
For the special case of the ground state in \eqref{dj} the degeneracy is given by the $N/2$ Catalan number $C_{N/2}$.
This degenerate ground state (macroscopic singlet state) is particularly interesting, e.g. for magnetometry \cite{Magneto} and black hole entropy calculations \cite{BHent}. In the next two subsections its degeneracy is stated for some higher spin systems.

\subsection{Spin 1 particles}

\begin{figure}[htbp]
\begin{center}
\scalebox{1}{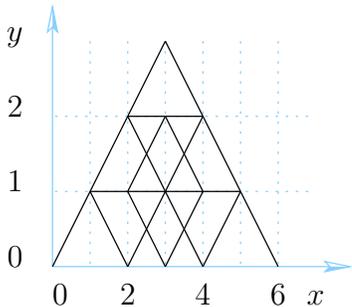}
\caption{Lattice paths illustrating $\text{mult}_{[0]} ([1]^{\otimes 6})=15$.}
\label{fig:spin1ex}
\end{center}
\end{figure}

Next we consider a system of $s=1$ particles.
When increasing $s$, the lattice paths become less standard due to additional constraints from \eqref{lattice_steps}.
For $s=1$ the paths can only reach integer values of $y$ and the allowed steps are
\begin{equation} \label{spin1}
\begin{aligned}
&\left\{
(1,1), (1,0), (1,-1)
\right\}, &y \geq 1,\\
&\left\{
(1,1)
\right\}, &y = 0.
\end{aligned}
\end{equation}
An example is given in Fig.\ \ref{fig:spin1ex}.
The paths with such constraints and $j=0$ are known in mathematical literature as Riordan paths (and their multiplicity as Riordan numbers \cite{Riordan1,Riordan2,Riordan3}. For $j \ge 0$ these are Riordan arrays \cite{Riordan4}, whose generating functions are known to be \cite{Riordan4}
\begin{equation}
\left(\frac{1+x-\sqrt{1-2x-3x^2}}{2x(1+x)},~\frac{1-x-\sqrt{1-2x-3x^2}}{2x}\right),
\end{equation}
from which the degeneracies $d_{1}(N,j)$ can be derived.
For instance, $d_{1}(N,0)$ is given by the Riordan numbers:
\begin{equation}
d_{1}(N,0)= \frac{1}{N+1}\sum_{k=1}^{N-1}\binom{N+1}{k}\binom{N-k-1}{k-1}.
\end{equation}
The recursion relation for the multiplicities \eqref{eq:m_recursion} becomes in this case
\begin{equation}
m_1(N,j) = \begin{cases}
1, &N=j=0,\\
\sum\limits_{k=|j-1|}^{\text{min}(j+1, N-1)}
m_1(N-1,k), &0 \leq j \leq N,\\
0, &\text{else}.
\end{cases}
\label{eq:m_recursion_s=1}
\end{equation}

\subsection{Higher spin particles}

For $s=\frac{3}{2}$ the allowed steps are
\begin{equation} \label{spin3/2}
\begin{aligned}
&\left\{
(1,\tfrac{3}{2}), (1,\tfrac{1}{2}), (1,-\tfrac{1}{2}), (1,-\tfrac{3}{2})
\right\}, &y \geq \tfrac{3}{2},\\
&\left\{
(1,\tfrac{3}{2}), (1,\tfrac{1}{2}), (1,-\tfrac{1}{2})
\right\}, &y =1,\\
&\left\{
(1,\tfrac{3}{2}), (1,\tfrac{1}{2})
\right\}, &y =\tfrac{1}{2},\\
&\left\{
(1,\tfrac{3}{2})
\right\}, &y =0.
\end{aligned}
\end{equation}
For example, the degeneracies for the case of $j=0$ are: \\
0, 1, 0, 4, 0, 34, 0, 364, 0, 4269, 0, 52844, 0, 679172, 0, 8976188, 0,
121223668. \\
As required, all the odd multiplicities vanish.
This sequence of integers (with or without the zeroes) is not known in mathematical literature but can explicitly solve the $s=3/2$ case as was done above for lower spin systems.

Continuing according to the same logic, the allowed steps for $s=2$ are
\begin{equation} \label{spin2}
\begin{aligned}
&\left\{
(1,2), (1,1), (1,0), (1,-1), (1,-2)
\right\}, &y \geq 2,\\
&\left\{
(1,2), (1,1), (1,0)
\right\}, &y =1,\\
&\left\{
(1,2)
\right\}, &y =0.
\end{aligned}
\end{equation}
For instance, the degeneracies for the case of $j=0$ are: \\
0, 1, 1, 5, 16, 65, 260, 1085, 4600, 19845, 86725, 383251, 1709566. \\
These degeneracies, and also the ones for $j>0$, are known in literature \cite{spintwo1,spintwo2}, but not in the context of 2D lattice paths.

We end this subsection by describing the degeneracies for the $s=3,~j=0$ case: \\
0, 1, 1, 7, 31, 175, 981, 5719, 33922, 204687, 1251460, 7737807,
48297536. \\
These again, are not known in mathematical literature.
We elaborate in the next section on the fraction of decideable states in all the above cases.

\section{3. The Fraction of Decidable States}

\begin{figure}[htbp]
\begin{center}
\begin{tikzpicture}[ spy using outlines={black,magnification=4, width=3cm, height=2cm,connect spies}]
    \node[anchor=south west,inner sep=0] (image) at (0,0){\includegraphics[width=0.47\textwidth]{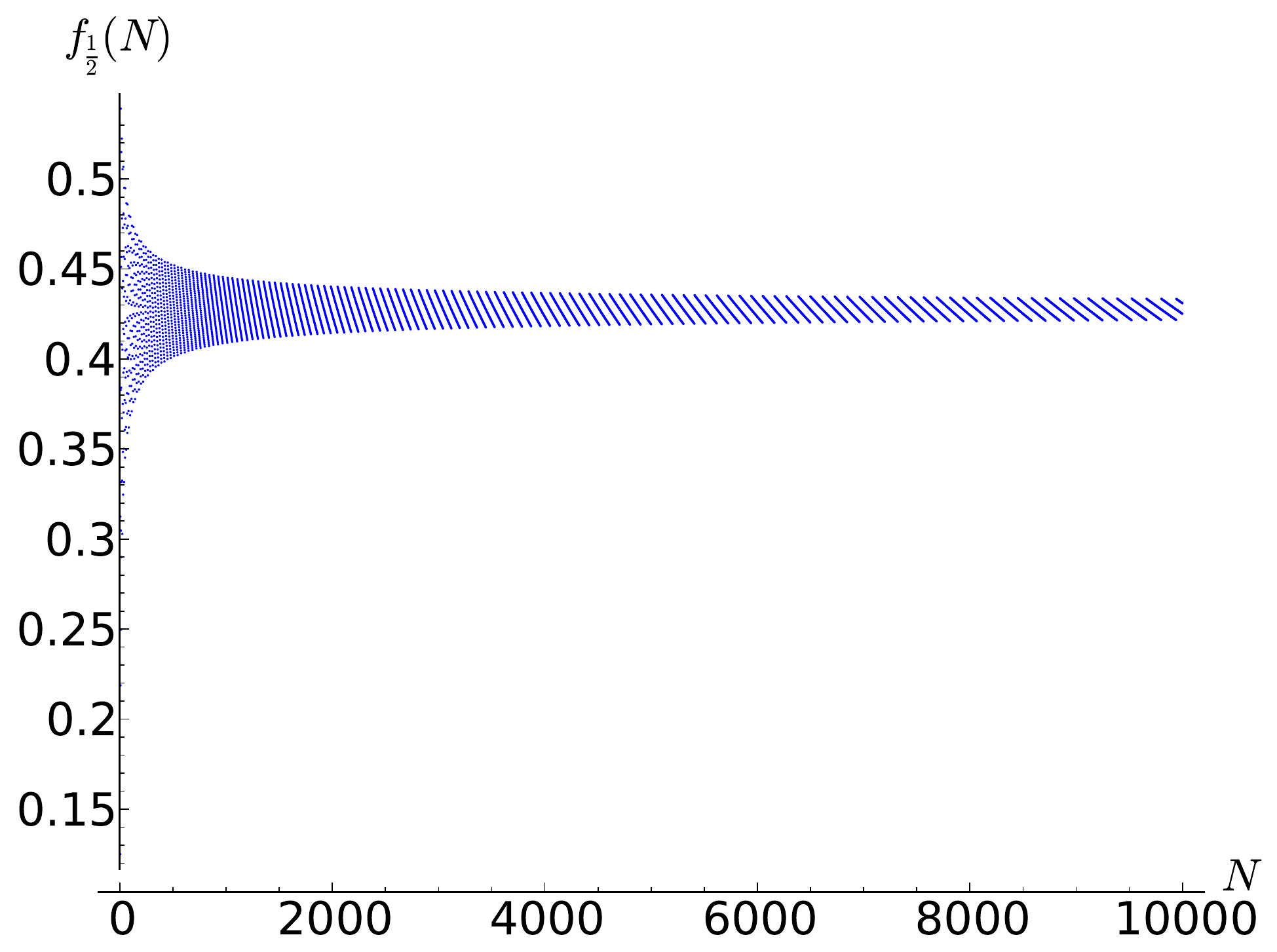}};
    \spy on ($0.67*(image.south east)+0.674*(image.north west)$) in node at ($0.67*(image.south east)+0.31*(image.north west)$);
\end{tikzpicture}
\caption{$f_{\frac{1}{2}}(N)$ for $N$ up to 10,000.}
\label{fig:spin12h}
\end{center}
\end{figure}

\begin{figure}[htbp]
\begin{center}
\begin{tikzpicture}[ spy using outlines={black,magnification=4, width=3cm, height=2cm,connect spies}]
    \node[anchor=south west,inner sep=0] (image) at (0,0){\includegraphics[width=0.47\textwidth]{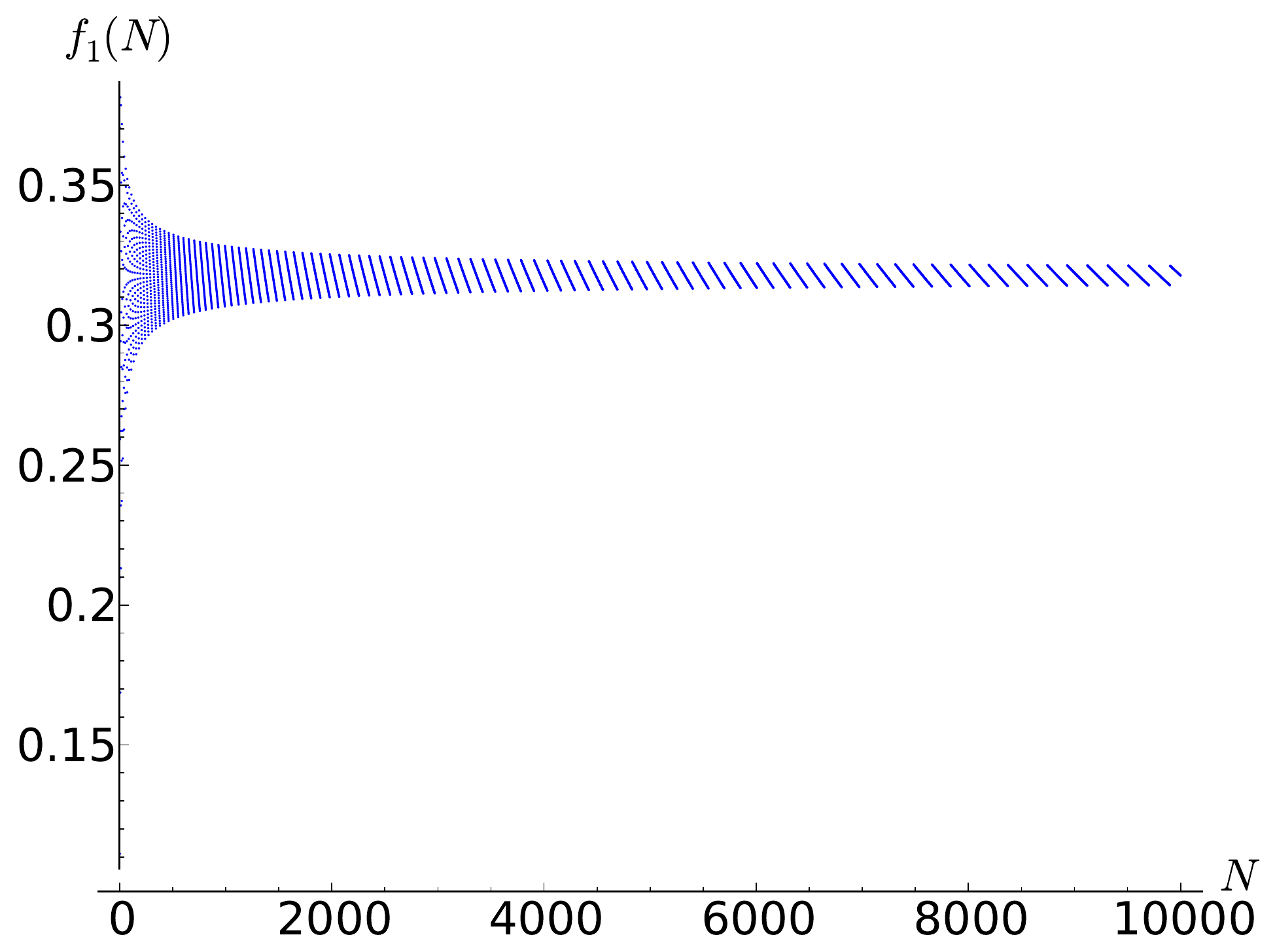}};
    \spy on ($0.67*(image.south east)+0.708*(image.north west)$) in node at ($0.67*(image.south east)+0.31*(image.north west)$);
\end{tikzpicture}
\caption{$f_1(N)$ for $N$ up to 10,000.}
\label{fig:spin1h}
\end{center}
\end{figure}

\begin{figure}[htbp]
\begin{center}
\includegraphics[width=0.47\textwidth]{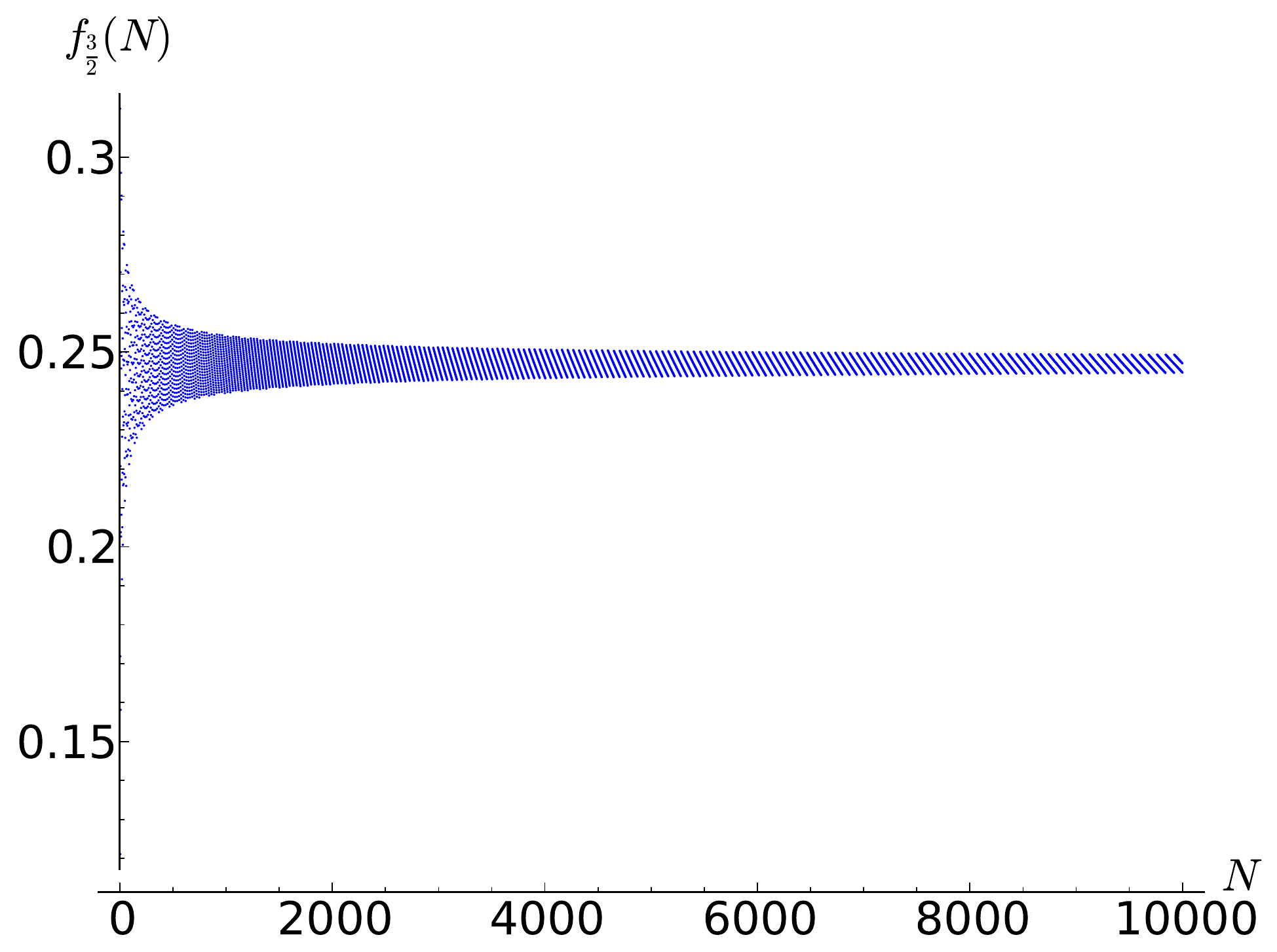}
\caption{$f_{\frac{3}{2}}(N)$ for $N$ up to 10,000.}
\label{fig:spin32h}
\end{center}
\end{figure}

\begin{figure}[htbp]
\begin{center}
\includegraphics[width=0.47\textwidth]{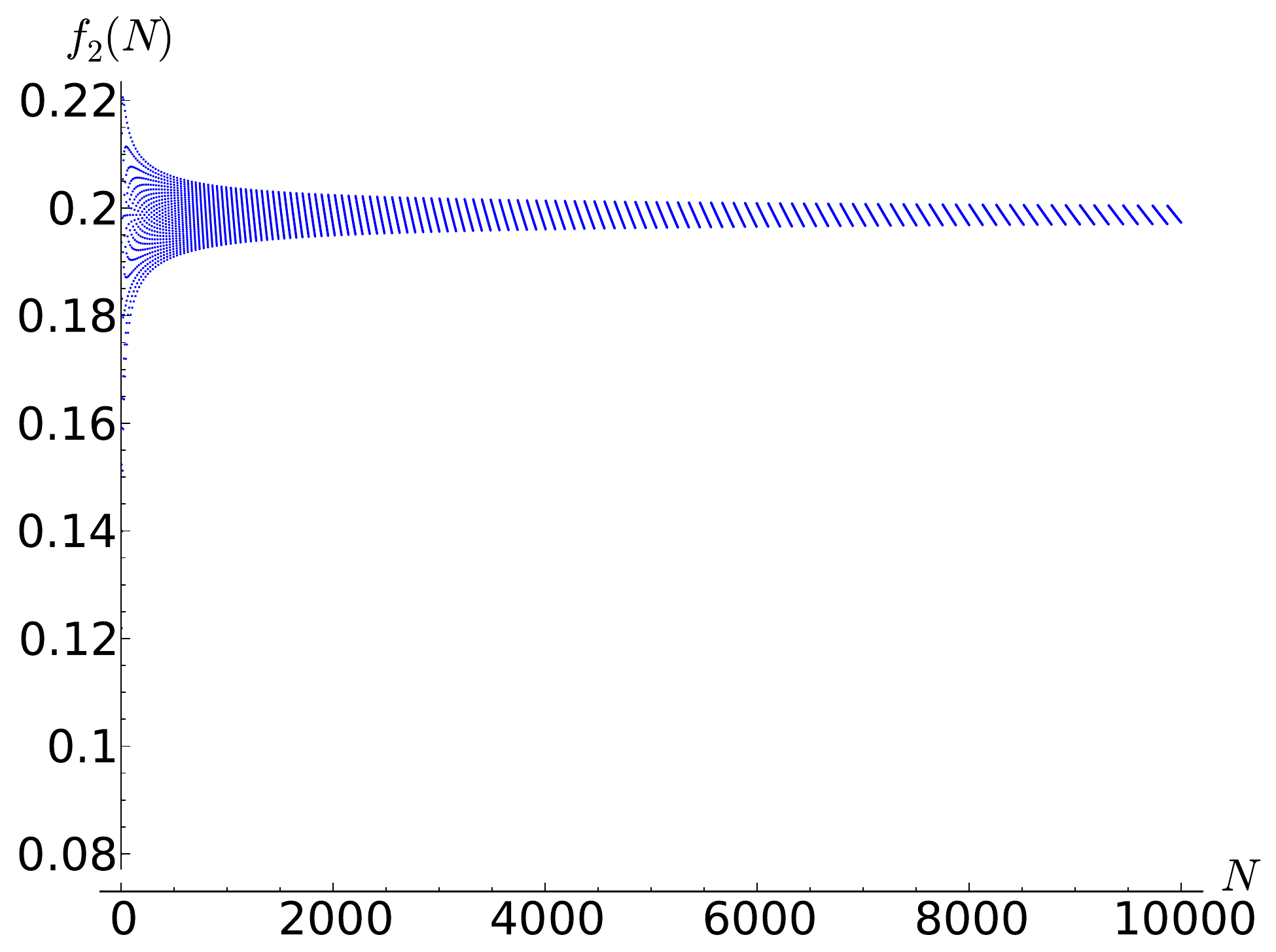}
\caption{$f_2(N)$ for $N$ up to 10,000.}
\label{fig:spin2h}
\end{center}
\end{figure}

\begin{figure}[htbp]
\begin{center}
\includegraphics[width=0.5\textwidth]{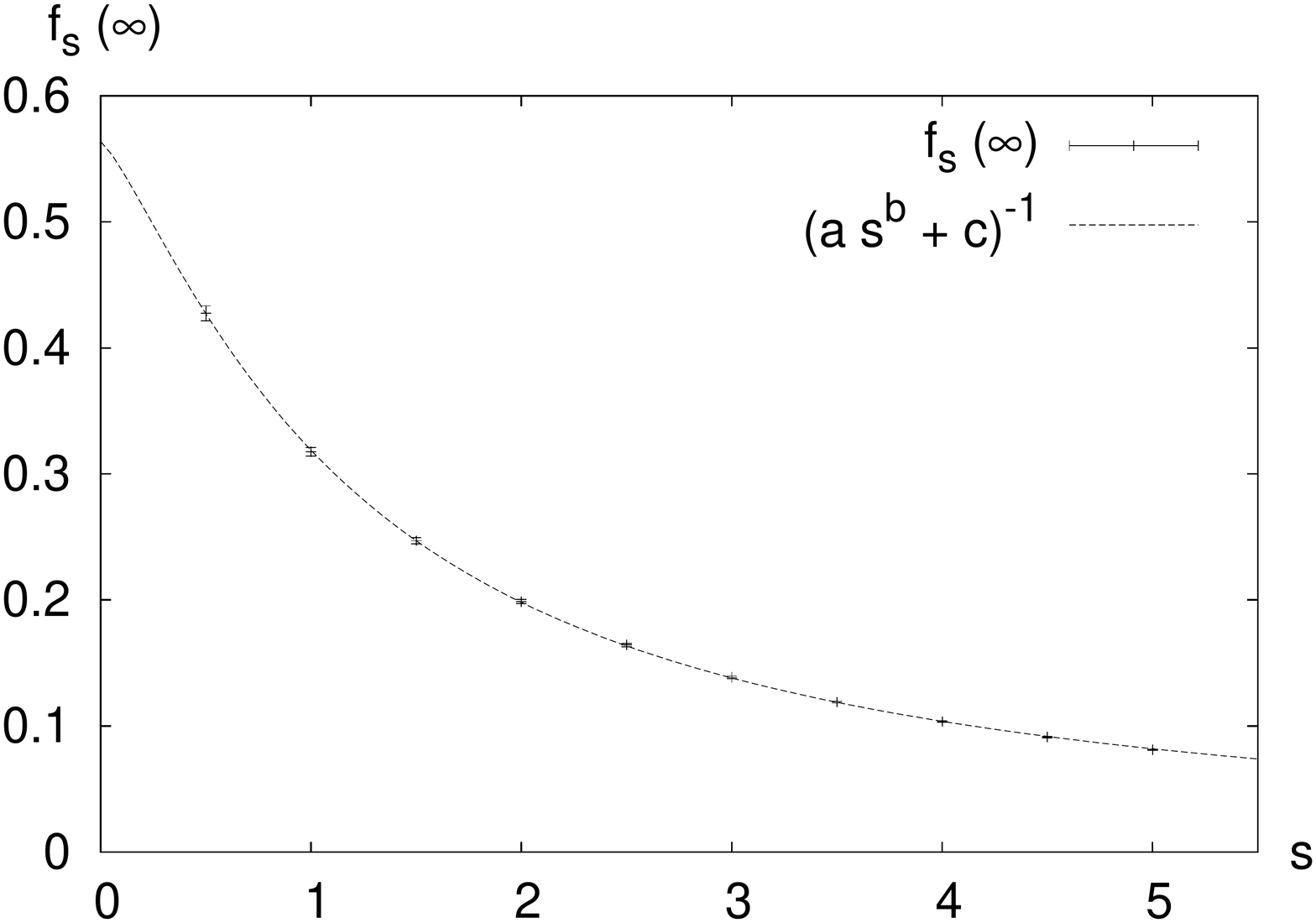}
\caption{$f_s(\infty)$ from Table\ \ref{tab:decidable_fractions} with fit (parameters given in \eqref{eq:fit_parameters}).}
\label{fig:f_infinity}
\end{center}
\end{figure}

To generalize \eqref{boundhlaf}, we employ the theory of entanglement detection with uncertainty relations \cite{UW,Hofmann}.
For every $N$ particle separable state it was shown that \cite{Hofmann,Totheq}:
\begin{equation}
(\Delta J_x)^2+(\Delta J_y)^2+(\Delta J_z)^2 \ge Ns.
\end{equation}
Therefore, if we define in a system of $N$ spin-$s$ particles the EW as the total magnetization \eqref{Wj}, then for separable states it is bounded from below by $Ns$:
\begin{equation}
\langle W \rangle \ge W_{sep}^{(s)}=Ns.
\end{equation}
All states with witness eigenvalues below this bound are entangled, so using the witness levels \eqref{Wj}
one finds the fraction of decidable states to be
\begin{equation}
f_s(N)=\frac{\sum\limits_{\{j \in \mathbb{N}_0/2 | j(j+1)<Ns\}} d_{s}(N,j)}{\sum\limits_{j \in \mathbb{N}_0/2} d_{s}(N,j)}.
\end{equation}
In Figures \ref{fig:spin12h}--\ref{fig:spin2h}  this fraction is plotted for $s=\frac{1}{2},1,\frac{3}{2},2$, respectively, where systems comprised of up to 10,000 spins were analyzed.
Computing the tensor products to such high orders was possible only owing to the recursion relation \eqref{eq:m_recursion}.
Interestingly, the points lie on curves that are constrained to a certain range. For half-integer $s$ the points for even and odd $N$ (the case of odd $N$ is studied here for the first time) lie on different curves, while for integer $s$ they lie on the same curves.
It seems that $f_s(N)$ converges for large $N$,
meaning that all points lie between two curves which monotonically approach the same constant
$f_{s}(\infty)$. A good approximation of these constants can be made based on the rightmost jump in $f_s(N)$ in a given graph.
For example, the last jump in $f_{\frac{1}{2}}(N)$ below $N=$10,000 is at
\begin{equation}
\begin{aligned}
f_{\frac{1}{2}}(9940) &\approx 0.42169,\\
f_{\frac{1}{2}}(9942) &\approx 0.43338,
\end{aligned}
\end{equation}
hence $f_{s}(\infty)$ must lie in between these values
\begin{equation}
\begin{aligned}
f_{\frac{1}{2}}(\infty) &= \frac{f_{\frac{1}{2}}(9942) + f_{\frac{1}{2}}(9940)}{2}
\pm \frac{f_{\frac{1}{2}}(9942) - f_{\frac{1}{2}}(9940)}{2}\\
&=0.4275 \pm 0.0058 .
\end{aligned}
\end{equation}
The corresponding values $f_{s}(\infty)$ for $s$ up to 5 are given in Table \ref{tab:decidable_fractions}.
They are plotted in Figure \ref{fig:f_infinity}, together with a fitted curve given by
\begin{equation}
f_{s}(\infty) \approx \frac{1}{a s^b + c},
\end{equation}
where
\begin{equation}
\label{eq:fit_parameters}
a = 1.36273, \qquad b = 1.26448, \qquad c = 1.7738.
\end{equation}
The sum of squared residuals of the fit is
\begin{equation}
\sum\limits_i \epsilon_i^2 = 3.6 \cdot 10^{-6}.
\end{equation}
The results of the graphs below suggest that for various spin systems of arbitrary size, there is a considerable amount of many-body entangled states that can be detected by the proposed EW. Even though this (quite natural) EW may not be the optimal one, it enables to identify a considerable sub-space of the multi-particle Hilbert space, comprised only of entangled states. These states, which are known to be more protected against decoherence
than other states corresponding to large values of $W$, can be used as a resource for measurement
based quantum computation and for quantum information storage \cite{Totheq}. For any $N$ and $s$, the $j=0$ states are especially important in that aspect, being ``decoherence free'' \cite{Paul}.

\begin{table}[htbp]
  \centering
  \begin{tabular}{|l|l l l|}
  \hline			
  $s$ & \multicolumn{3}{|c|}{$f_s(\infty)$} \\ \hline
  $\frac{1}{2}$ & $0.4275$ & $\pm$ & $0.0058$ \\
  $1$ & $0.3177$ & $\pm$ & $0.0035$ \\
  $\frac{3}{2}$ & $0.2470$ & $\pm$ & $0.0023$ \\
  $2$ & $0.1987$ & $\pm$ & $0.0017$ \\
  $\frac{5}{2}$ & $0.1642$ & $\pm$ & $0.0013$ \\
  $3$ & $0.1386$ & $\pm$ & $0.0010$ \\
  $\frac{7}{2}$ & $0.11897$ & $\pm$ & $0.00082$ \\
  $4$ & $0.10356$ & $\pm$ & $0.00068$ \\
  $\frac{9}{2}$ & $0.09119$ & $\pm$ & $0.00056$ \\
  $5$ & $0.08110$ & $\pm$ & $0.00049$ \\
  \hline
  \end{tabular}
  \caption{Approximate values for $f_s(\infty)$ based on the last jump appearing below $N=$10,000.}
  \label{tab:decidable_fractions}
\end{table}

\section{4. Conclusions}
We have introduced in this work a tripartite analogy between the degeneracy of witness eigenstates, tensor products of $SO(3)$ representations and classical lattice walks with special constraints. Furthermore, we found that the solution to the above problems is given by generalized Catalan numbers. This analogy enabled us to construct a ``sterile'' entanglement witness for arbitrary spin systems, which marginally changes them upon measuring highly entangled states. Being an important resource for various quantum information processing tasks, we have derived the fraction of decideable states for such a witness and examined its dependency on the spin $s$ and the number of particles $N$. It was found to be a decreasing function in $s$ and an asymptotically constant function in $N$. \\

{\bf Acknowledgements.}
We would like to thank Yakir Aharonov, Roy Ben-Israel, Wojciech Samotij and G$\acute{\text{e}}$za T$\acute{\text{o}}$th for helpful comments and discussions. E.C. was supported by Israel Science Foundation Grant No. 1311/14 and ERC AdG NLST.
T.H. received funding from the grant CERN/FIS-NUC/0045/2015 and was supported by the [European Union] 7th Framework Programme (Marie Curie Actions) under grant agreements No 269217 (UNIFY) and 317089 (GATIS) and by the German Science Foundation (DFG) within the Collaborative Research Center 676 ``Particles, Strings and the Early Universe''.
\emph{Centro de Fisica do Porto} is partially funded by the Foundation for Science and Technology of Portugal (FCT).
N.I. was supported in part by the I-CORE Program of the Planning and Budgeting Committee and the Israel Science Foundation (Center No. 1937/12), and by a center of excellence supported by the Israel Science Foundation
(grant number 1989/14).

\end{document}